\documentclass[iop]{emulateapj}
\usepackage{latexsym}
\usepackage{dcolumn}
\usepackage{bm}
\usepackage{amsmath}
\usepackage{natbib}
\input{epsf}
\newcommand{\be}{\begin{equation}}
\newcommand{\ee}{\end{equation}}
\newcommand{\ba}{\begin{eqnarray}}
\newcommand{\ea}{\end{eqnarray}}
\newcommand{\bi}{\begin{itemize}}
\newcommand{\ei}{\end{itemize}}
\newcommand{\bfi}{\begin{figure}[t]
\epsfxsize=9cm
\epsffile}
\newcommand{\bfig}{\begin{figure*}[t]
\epsfxsize=15cm
\epsffile}
\newcommand{\efi}{\end{figure}}
\newcommand{\efig}{\end{figure*}}

\begin{document}
\title{Weak lensing power spectrum reconstruction by counting
  galaxies.-- II: Improving the ABS method with the shift parameter}
\author{Pengjie
  Zhang$^{1,2,3,4}$, Xinjuan Yang$^5$, Jun Zhang$^{1,4}$, Yu Yu$^{1,4}$ \\
$^1$Department of Astronomy, School of Physics and Astronomy, Shanghai Jiao Tong University, 955
Jianchuan road, Shanghai, 200240\\ 
$^2$ IFSA Collaborative Innovation Center, Shanghai Jiao Tong
University, Shanghai 200240, China\\
$^3$ Tsung-Dao Lee Institute, Shanghai 200240, China\\
$^4$ Shanghai Key Laboratory for Particle Physics and Cosmology\\
$^5$ Institute for Advanced Physics \& Mathematics, Zhejiang
University of Technology, Hangzhou, 310032, China
}
\email[Email me at: ]{zhangpj@sjtu.edu.cn}
\begin{abstract}
In paper I of this series (Yang et al. 2017, ApJ), we proposed an
analytical method of blind separation ({\bf ABS}) to extract the  cosmic
magnification signal in galaxy number distribution and reconstruct the
weak lensing power spectrum. Here we report a new version of the
ABS method, with significantly improved performance. This version is characterized by a shift parameter
$\mathcal{S}$, with the special case of $\mathcal{S}=0$ corresponding
to the original ABS method. We have tested this new version, compared
to the previous one, and  confirmed its supreme performance in
all investigated situations. Therefore it supersedes the previous
version. The proof of concept studies presented in this paper
demonstrate that it may enable surveys such as  LSST and 
SKA to reconstruct the lensing  power spectrum at $z\simeq 1$ with
$1\%$ accuracy. We will test with more realistic simulations to verify
its applicability in real data.   
\end{abstract}
\maketitle

\section{introduction}
Gravitational lensing not only distorts galaxy images and induces the
cosmic shear effect, but also
changes the spatial distribution of galaxies and induces the cosmic
magnification effect
\citep{1995A&A...298..661B,2001PhR...340..291B}. This cosmic
magnification effect provides a way of lensing
measurement alternative to the cosmic shear. However, despite many
appealing advantages, its extraction from the 
overwhelming intrinsic fluctuations of galaxy spatial distribution is
highly challenging. \citet{Zhang05} pointed out
that the two are in principle separable in the flux space,  due to their different
dependences on galaxy flux. \citet{YangXJ11,YangXJ15} developed
algorithms of implementing this
idea at map and power spectrum level, respectively. Both works
identified the galaxy stochasticity as the major challenge in this
exercise.  In principle we can fit the galaxy clustering and the
lensing power spectrum simultaneously. However, this induces a model
dependence on the galaxy clustering, in particular on its stochastic
part. Furthermore,  to achieve percent level accuracy,  many parameters (such as the scale and flux
dependence of deterministic and stochastic bias) should be involved in
the fitting.  It can be computationally 
challenging and numerically unstable.

 These problems can be overcome by the ABS method developed by two of
the authors \citep{ABS} in the context of CMB foreground removal. \citet{YangXJ17} applied the ABS method and
demonstrated its applicability. The ABS method  does
not rely on assumptions of  galaxy intrinsic clustering, providing a
blind separation of cosmic magnification from galaxy intrinsic
clustering.  It is computationally fast, since only a few linear
algebra operations are needed. The ABS method is exact,  when measurement
errors (shot noise) in the galaxy clustering measurement is
negligible. However, when shot noise exceeds
certain level, it may become  biased and numerically
instable. Similar problem exists in the case of CMB. Recently we have
constructed a new version of the ABS method 
(\citet{ABS}, version 2), with a newly introduced shift parameter
$\mathcal{S}$. It is also based on exact solutions as the original
version, which corresponds to the limit of $\mathcal{S}=0$. The new version solves
the problem in CMB foreground removal. A natural step is to
apply this new version to cosmic magnification. As will be shown in
this paper, the improvement is significant. The ABS reconstructed
lensing power spectrum remains unbiased and numerically stable even
for cases of large noise.   Furthermore, despite of
being two independent systems, the same $\mathcal{S}$ works for both
CMB and cosmic magnification and therefore no fine tuning is
required. We conclude that it should supercede the previous
one, and report this new version in the paper II of this series. 
\section{The new version of ABS with the shift parameter $\mathcal{S}$}
\label{sec:ABS}
Here we briefly summarize the equation that ABS solves in the context
of cosmic magnification. For detailes, please refer to paper I.  The ABS method solves the following equation,
\be
\label{eqn:CLobs}
C^{\rm obs}_{ij}=C^L_{ij}+\delta C_{ij}^{\rm shot}\ .
\ee
$C^{\rm obs}_{ij}$ is the cross power spectrum between the galaxy
distribution of the $i$-th and $j$-th flux bins. In this expression, the ensemble average
of shot noise power spectrum has been subtracted from the diagonal
elements ($i=j$). What left is the residual due to statistical
fluctuation, $\delta C_{ij}^{\rm shot}$ ($\langle \delta C_{ij}^{\rm shot}\rangle$). Without loss of generality, we choose flux bins such
that different bins have identical error ($\sigma^2_{\rm shot}\equiv
\langle (\delta C_{ii}^{\rm shot})^2\rangle$). $C^L_{ij}(\ell)$ is
the cross power spectrum between galaxies in the 
$i$-th and $j$-th flux bins, in the limit of negligible shot noise.
This astrophysical signal has three contributions, the intrinsic galaxy
auto power spectrum $C_{ij}^g$, the cosmic magnification
auto power spectrum, and the cross power spectrum between the galaxy
intrinsic clustering and cosmic magnification (paper I).  As shown in paper I,
it can be formulated into the following form,  
\ba
\label{eqn:CL}
C^L_{ij}(\ell)=g_ig_j\tilde{C}_\kappa+\tilde{C}^g_{ij}\ .
\ea
Here
\ba
\label{eqn:transformation}
\tilde{C}_\kappa &\equiv& C_\kappa(1-r^2_{m\kappa}) \ .
\ea
$C_\kappa$ is the lensing power spectrum. $r_{m\kappa}$ is the cross
correlation coefficient between lensing and the matter distribution over the
redshift range of source galaxies. The prefactor $g(F)$ is determined by $n(F)$, the average number of galaxies per flux
interval. For a narrow flux bin, $g=2(\alpha-1)$ where $\alpha\equiv -d\ln n/d\ln
F-1$. $\tilde{C}^g_{ij}$ is basically the galaxy intrinsic
clustering whose exact definition is given in paper I. 

\bfi{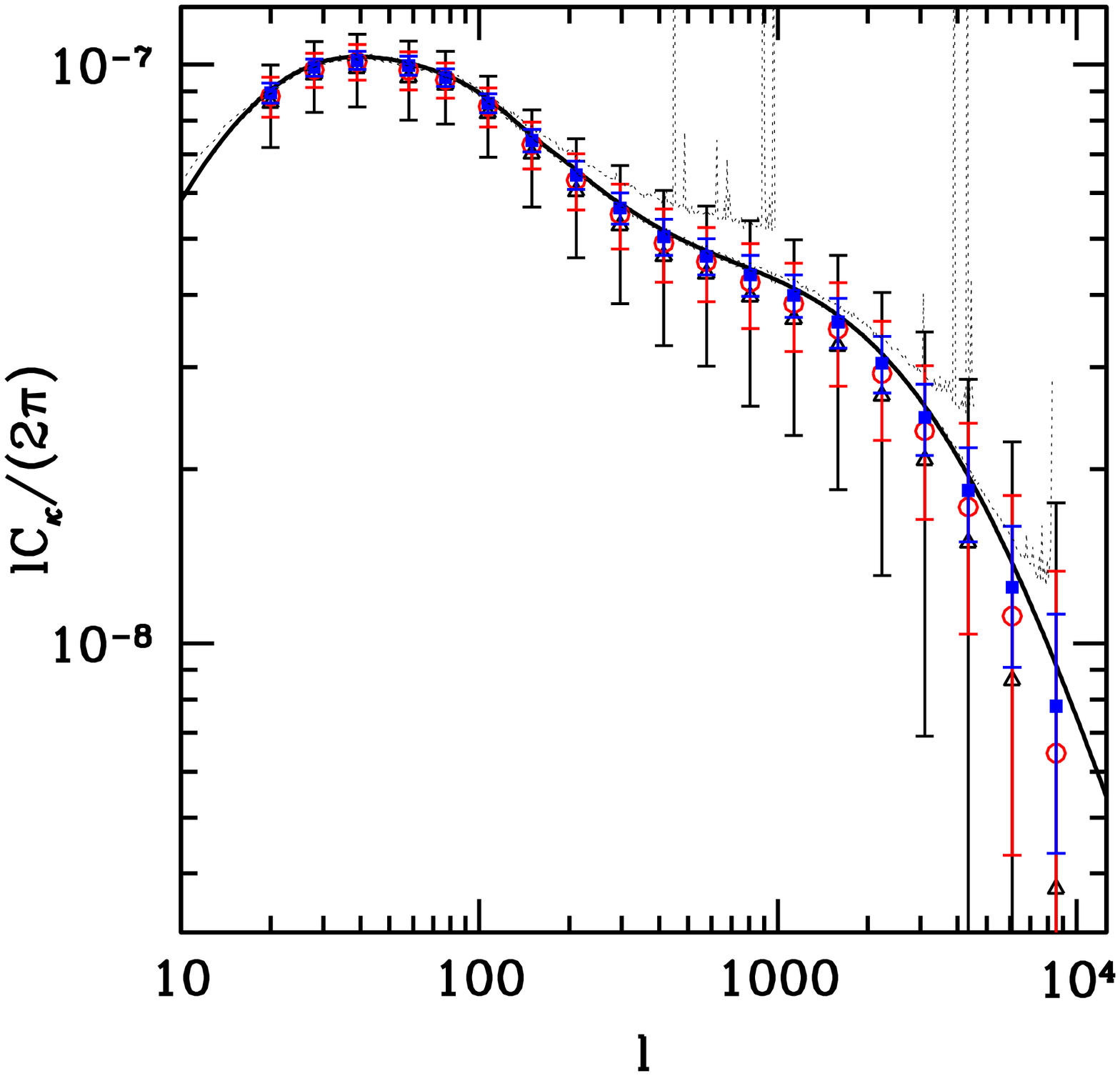}
\caption{The improved ABS method with the shift parameter
  $\mathcal{S}=20\sigma_{\rm shot}$.  Errorbars are estimated using
  $1000$ realizations of shot noise. The 3 sets of data points with
  errorbars correspond to survey specifications S1, S2 and S3
  respectively. Triangle data (black) points with largest errorbars
  correspond to S1 with smallest galaxy number. Square data (blue)
  points with smallest errorbars correspond to S3 with highest galaxy
  number. Open circle data points correspond to S2. For comparison,
  we overplot the results (dot lines) of the original ABS method
  ($\mathcal{S}=0$). The previous method breaks when shot noise exceeds
  certain threshold. In contrast,  the new method remains unbiased
  and numerically stable.  \label{fig:kappaA}}
\efi

\bfi{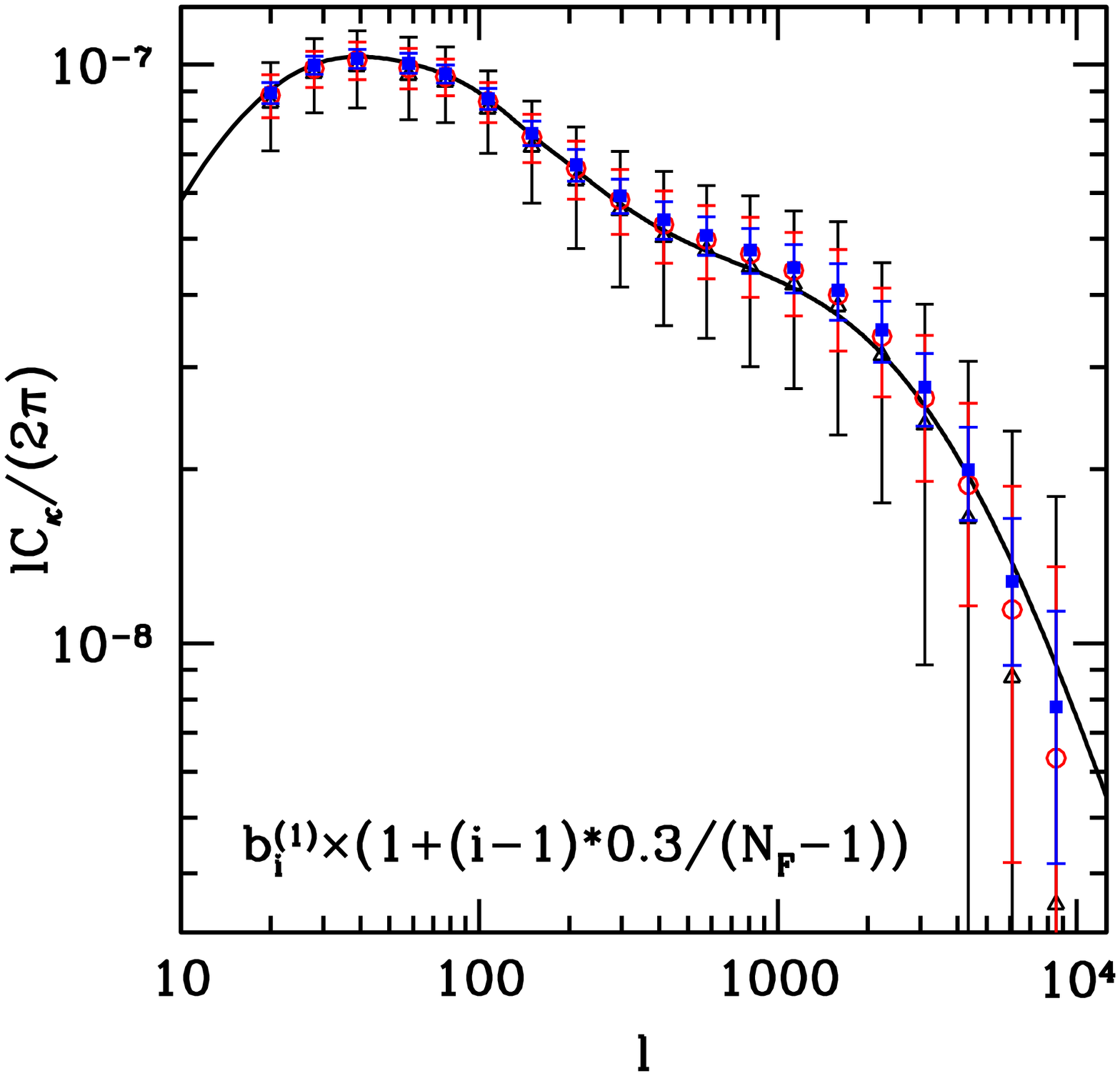}
\caption{Similar to Fig. \ref{fig:kappaA}, but for case B where the
  shape of ${\bf
    b}^{(1)}$ is changed by $30\%$. \label{fig:kappaB} }
\efi
The ABS method solves Eq. \ref{eqn:CLobs} for $\tilde{C}_\kappa$,
based on the fact that $g(F)$ is an observable, and its flux
dependence differs from that of the galaxy intrinsic
clustering. When the number of flux
bins is larger than the number of eigenmodes 
 of $C^g_{ij}$, the solution to $\tilde{C}_\kappa$ is unique and
unbiased. Hereafter we will work under this condition.  Following the new version  of the ABS method (\citet{ABS},
version 2), the
estimator of $\tilde{C}_\kappa$ is 
\ba
\label{eqn:ABS}
\hat{\tilde{C}}_{\kappa}=\left(\sum_{\lambda_\mu>\lambda_{\rm
      cut}\sigma_{\rm shot}}
  G_\mu^2\lambda_\mu^{-1}\right)^{-1}_{C^{\rm obs}_{ij}+g_ig_j\mathcal{S}}-\mathcal{S}\ . 
\ea
Here, $\mathcal{S}$ is the shift parameter of any value. $\lambda_\mu$
is the $\mu$-th eigenvalue of the matrix $C^{\rm
  obs}_{ij}+g_ig_j\mathcal{S}$. The corresponding eigenvector is ${\bf
  E}^{(\mu)}$ and $G_\mu\equiv {\bf E}^{(\mu)}\cdot {\bf g}$. With the
presence of noise, some eigenmodes may be heavily polluted or even
completely unphysical. We have to exclude them. Therefore we add a cut and only use
eigenmodes with eigenvalue above the threshold $\lambda_{\rm
  cut}\sigma_{\rm shot}$.  

\bfi{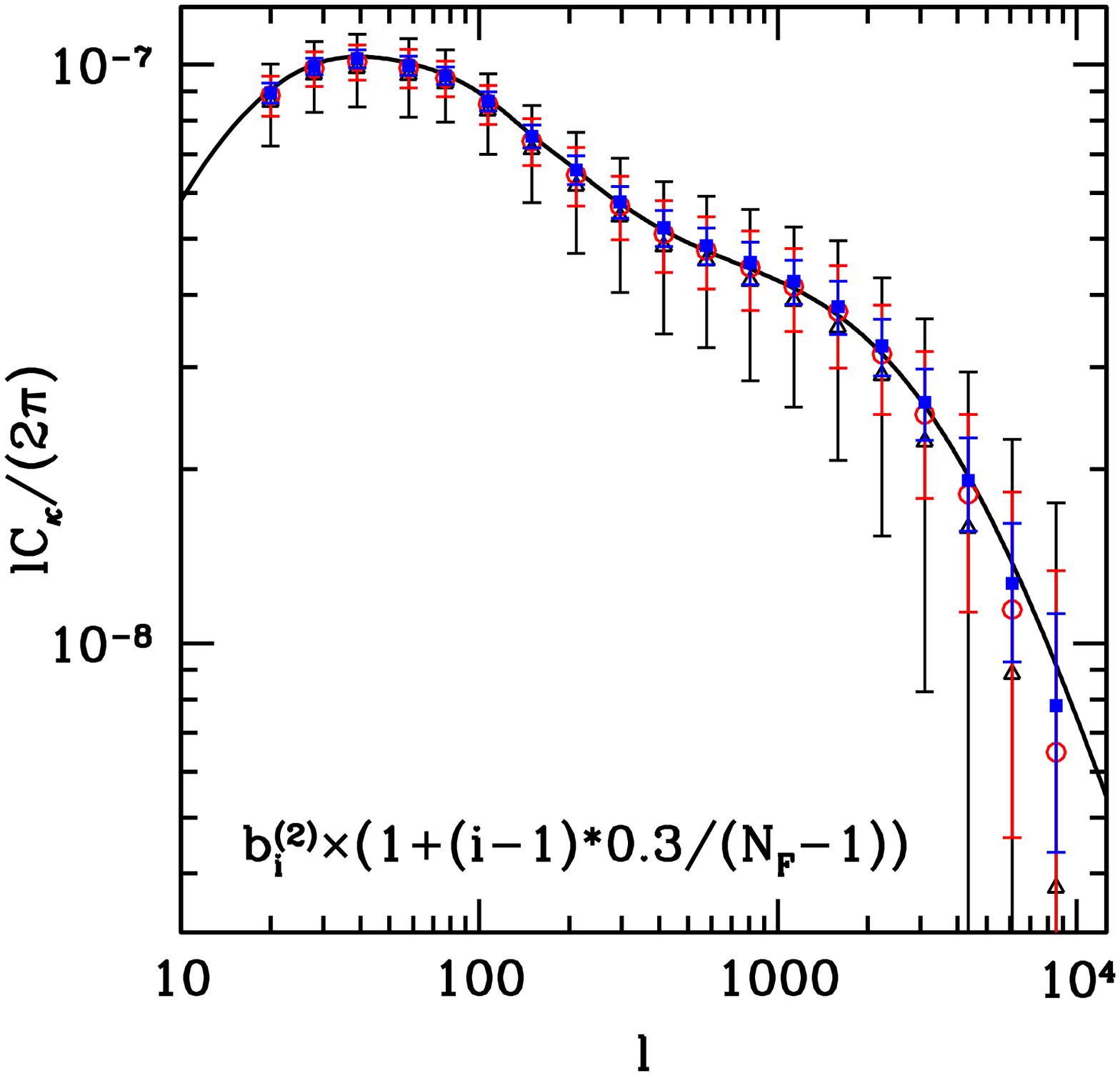}
\caption{Similar to Fig. \ref{fig:kappaA}, but for case C where the
  shape of ${\bf
    b}^{(2)}$ is changed by $30\%$. \label{fig:kappaC} }
\efi

 Eq. \ref{eqn:ABS} is exact when $\delta C^{\rm shot}_{ij}=0$. In this
 ideal case, the choice of $\mathcal{S}$ is irrelevant. However, with
 the presence of measurement error, its choice indeed makes
 difference.  Therefore, $\mathcal{S}$, despite its dimension the same
 as $\tilde{C}_\kappa$, is essentially a regularization parameter associated with
 the matrix operation. 
The original ABS method used in paper I is the special case of
$\mathcal{S}=0$.  However, with the presence of residual shot noise, such version
can not pass the null test. In this case,  the true signal is zero,
while the value returned by the ABS method is always postive. Appropriate choice of $\mathcal{S}$ can solve this
problem. Since the first term at the right hand side of
Eq. 4 is always positive,  $\mathcal{S}$  must be positive in order to pass the null
test.  Furthermore, it has to satisfy $\mathcal{S}\gg \sigma_{\rm
  shot}$.  Meanwhile, a positive $\mathcal{S}$ improves the
numerical stability. When $\mathcal{S}\gg \sigma_{\rm
  shot}$, it also  passes the convergence test.  In the context of CMB B-mode foreground removal,
we find that $\mathcal{S}=20\sigma^{\rm inst}_{\mathcal{D}}$ is 
a good choice (\citet{ABS}, version 2).   It is self-determined within the data through the
convergence test and the same choice of $\mathcal{S}$ automatically passes the null test. In this paper, we
will adopt the same shift parameter $\mathcal{S}=20\sigma_{\rm shot}$,
along with the same cut $\lambda_{\rm cut}=1/2$.  These values may not be the
optimal choice for lensing reconstruction. However, to avoid fine
tunings and uncertainties associated with them, we will fix
$\mathcal{S}/\sigma_{\rm shot}=20$ and $\lambda_{\rm cut}=1/2$.
Later we will find that the performance of the ABS method with these
fixed values is already excellent. Therefore fine tuning in
$\mathcal{S}$ and $\lambda_{\rm cut}$ is not required. 

\bfi{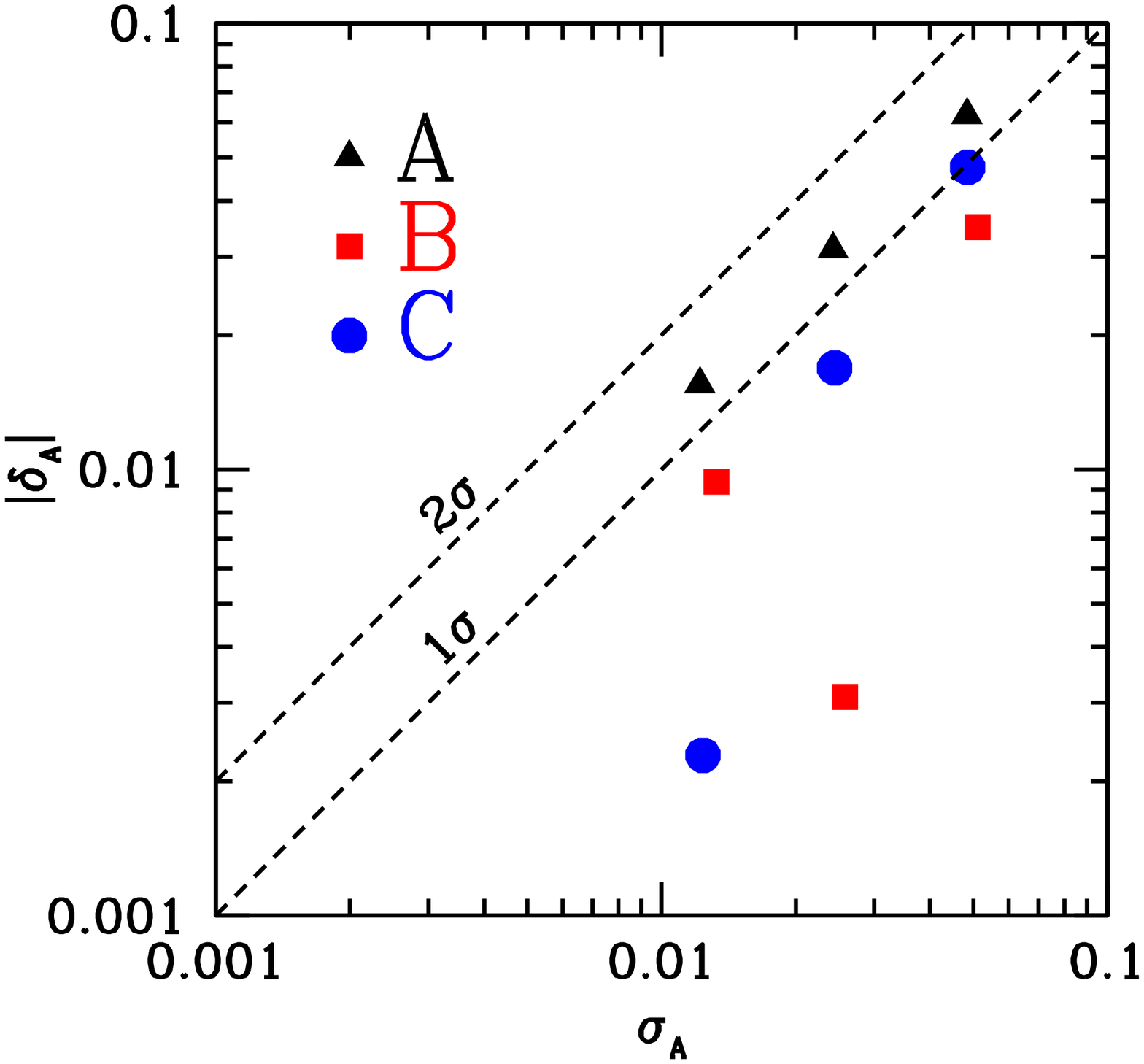}
\caption{The systematic error $\delta_A$ and statistical error $\sigma_A$  in the overall
  amplitude of the reconstructed
  lensing power spectrum. Smaller
  $\sigma_A$ corresponds to smaller galaxy number density. Data points
  of triangle, square and circle are for case A, B \& C of galaxy
  intrinsic clustering, respectively.  No systematic error above
  $1.5\sigma$ confidence level is detected.  \label{fig:sys}}
\efi

\section{Test results}
We follow the same set up of paper I to test the new ABS method. We
adopt 5 flux bins for galaxies in $0.8<z<1.2$. We include both the
deterministic and quadratic bias of galaxies. Along with the survey
specifications detailed in paper I, this fixes $C^L_{ij}$. For each fixed $C^L_{ij}$,  we
generate $1000$ realizations of $\delta C^{\rm shot}_{ij}$, assuming a
Gaussian distribution with zero mean and r.m.s. $\sigma_{\rm
  shot}$. $\sigma_{\rm shot}$ is evaluated adopting the sky coverage $10^4$
deg$^2$ and the total number of galaxies $N_{\rm tot}$.  We adopt the same
survey specifications (S1, S2 \& S3) as paper I.  S1, with $N_{\rm tot}=10^9$, resembles
a stage IV dark energy survey such as LSST or SKA.  S2 has $N_{\rm
  tot}=5\times 10^8$ and S3 has $N_{\rm tot}=2.5\times 10^8$. Paper I showed that the
performance of ABS depends on both the survey specifications and
properties of the galaxy intrinsic clustering. We test different cases
of galaxy intrinsic clustering.  Case A is the fiducial one,  with the linear bias ${\bf b}^{(1)}$ and quadratic
bias ${\bf b}^{(2)}$ specified in Fig. 2 of paper I. Case B changes
the shape of ${\bf b}^{(1)}$ from faint end to positive end by
$30\%$. Case C changes
the shape of ${\bf b}^{(2)}$ from faint end to positive end by
$30\%$. For case B and C, the previous ABS method shows visible systematic error for some
$\ell$ (Fig. 9 \& 10, paper I). Therefore
we choose to test the new ABS method using them.

Fig. \ref{fig:kappaA} shows the test result for galaxy intrinsic
clustering case A. Errorbars are estimated using $1000$ realizations
of shot noise ($\delta^{\rm shot}_{ij}$). For the survey specification of lowest galaxy
number density (S1), the previous ABS method breaks at $\ell\sim 400$, where
significant systematic error and numerical instability
develop. Increasing the galaxy number density pushes the scale of failure to
higher $\ell$, but the problem remains. In
contrast, the new ABS method solves this problem. It remains numerically
stable even for large shot noise. It remains unbiased at all scales of
interest. 

\bfi{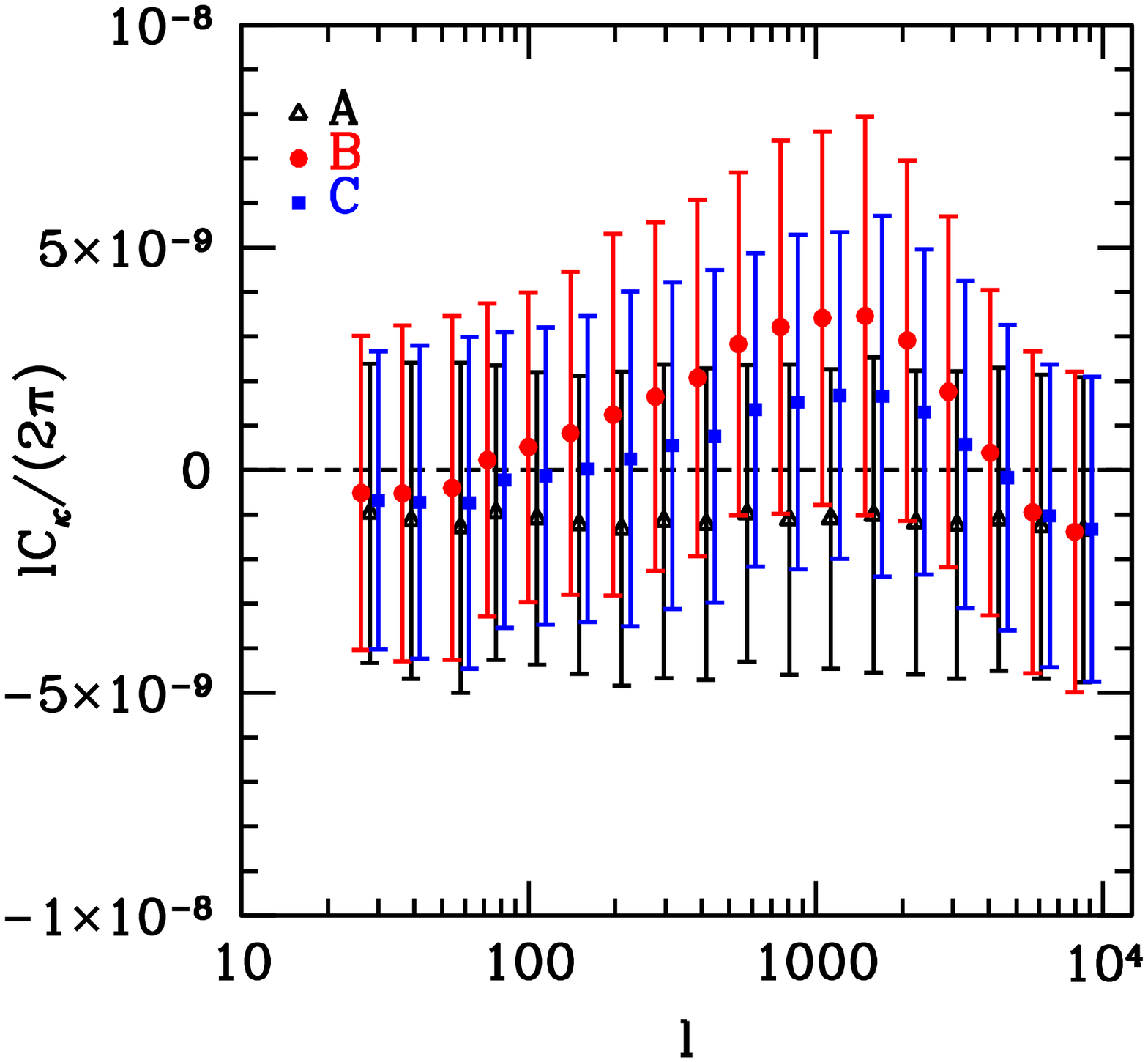}
\caption{The null test. We set the input lensing signal as zero and
  check if the output is consistent with zero. All $3\times 3$ cases
  investigated pass the null test. For clarity, we only show the 
  results for the survey specification ${\bf S}$1, and shift the data
  points of set A,B and C horizontally. \label{fig:null}}
\efi

Tests on case B (Fig. \ref{fig:kappaB}) and C (Fig. \ref{fig:kappaC})
show similar improvement on systematic bias and numerical
stability. These tests imply that the improvement on systematic bias and numerical
stability is general, not limiting to special cases of galaxy
intrinsic clustering. 

We compress the above results into the statistical error $\sigma_A$
and systematic error $\delta_A$ in the overall amplitude of the
lensing power spectrum. Fig. \ref{fig:sys} shows the results for three
cases of galaxy number density and three cases of galaxy intrinsic
clustering. It shows that the lensing reconstruction by the new ABS
method is statistically unbiased at $\sim 1\sigma$ level. 

This conclusion is further consolidated by the null test. We set the
lensing signal as zero and check whether the ABS output is consistent
with zero. For all $3\times 3$ cases above (3 cases of bias by 3
cases of shot noise), the ABS method passes the
null test (Fig. \ref{fig:null}). 

Nevertheless, there are many issues for further investigation when applying the method to
real data. We need to deal with survey complexities such as
masks, photometric errors and photo-z errors. (1) For masks, we may
need to measure the angular correlation functions 
first. By adopting  estimators such
as the Landy-Szalay estimator \citep{1993ApJ...412...64L},  the
measured  correlation functions can be free of masks. We can then
Fourier transform them to obtain the power spectra and apply the ABS
method to reconstruct the lensing power spectrum. Alternatively, we
can directly apply the ABS method to the measured correlation
functions and reconstruct the lensing correlation function.  (2) We have discussed the photometry calibration error in paper
I. The ABS method is applicable with the existence of photometry
calibration error, but the r.m.s. must be known. (3) Photo-z errors lead
to inaccurate determination of $g$ and therefore  impact the
reconstruction.  However, since the photo-z error for future surveys
such as LSST is smaller than the adopted photo-z bin size $\Delta
z^P=0.4$, this effect is expected to be sub-dominant. We also need
more realistic input of galaxy intrinsic clustering, whose
stochasticities can go beyond the adopted model of quadratic bias.  We are using
N-body simulations to generate galaxy mocks with these complexities
included, and test the ABS method  in a more robust and more realistic
way. 

\section{Conclusions}
We report a new version of the ABS method in lensing reconstruction by
counting galaxies. With a shift parameter
$\mathcal{S}$ about 20 times the measurement noise, the new ABS method
significantly improves the systematic bias and numerical
stability. For all cases investigated, the new ABS method remains statistically
unbiased and numerically stable. Therefore it supersedes the previous
version \citep{YangXJ17}. When applying to future surveys such as
LSST and SKA, it is promising to reconstruct the $z\sim 1$ lensing
power spectrum with $1\%$ accuracy. In future works, we will apply
this new version of ABS to simulated data and eventually to real
data.  Both in paper I and the current paper we work on the power spectrum measurement. The
ABS method also applies to the correlation functions.  We just need to
replace the  matrix of power spectra in Eq. 4 with the corresponding
matrix of cross correlation functions.

\section*{Acknowledgments}
This work was supported by the National Science Foundation of China
(11621303, 11433001, 11653003, 11320101002,11603019, 11403071, 11475148),  National
Basic Research Program of China (2015CB85701) and Zhejiang
province foundation for young researchers (LQ15A030001).

\bibliographystyle{apj}
\bibliography{mybib}

\begin{thebibliography}{}
\expandafter\ifx\csname natexlab\endcsname\relax\def\natexlab#1{#1}\fi

\bibitem[{{Bartelmann}(1995)}]{1995A&A...298..661B}
{Bartelmann}, M. 1995, \aap, 298, 661

\bibitem[{{Bartelmann} \& {Schneider}(2001)}]{2001PhR...340..291B}
{Bartelmann}, M., \& {Schneider}, P. 2001, \physrep, 340, 291

\bibitem[{{Landy} \& {Szalay}(1993)}]{1993ApJ...412...64L}
{Landy}, S.~D., \& {Szalay}, A.~S. 1993, \apj, 412, 64

\bibitem[{{Yang} {et~al.}(2017){Yang}, {Zhang}, {Yu}, \& {Zhang}}]{YangXJ17}
{Yang}, X., {Zhang}, J., {Yu}, Y., \& {Zhang}, P. 2017, \apj, 845, 174

\bibitem[{{Yang} \& {Zhang}(2011)}]{YangXJ11}
{Yang}, X., \& {Zhang}, P. 2011, \mnras, 415, 3485

\bibitem[{{Yang} {et~al.}(2015){Yang}, {Zhang}, {Zhang}, \& {Yu}}]{YangXJ15}
{Yang}, X., {Zhang}, P., {Zhang}, J., \& {Yu}, Y. 2015, \mnras, 447, 345

\bibitem[{{Zhang} \& {Pen}(2005)}]{Zhang05}
{Zhang}, P., \& {Pen}, U.-L. 2005, Physical Review Letters, 95, 241302

\bibitem[{{Zhang} {et~al.}(2016){Zhang}, {Zhang}, \& {Zhang}}]{ABS}
{Zhang}, P., {Zhang}, J., \& {Zhang}, L. 2016, ArXiv e-prints, arXiv:1608.03707

\end{thebibliography}

\end{document}